\documentclass[conference]{IEEEtran}
\IEEEoverridecommandlockouts
\usepackage{cite}
\ifCLASSINFOpdf
\usepackage[pdftex]{graphicx}
\else
\fi
\usepackage{amsmath}
\usepackage{amsfonts}
\usepackage{array}

\ifCLASSOPTIONcompsoc
\usepackage[caption=false,font=normalsize,labelfont=sf,textfont=sf]{subfig}
\else
\usepackage[caption=false,font=footnotesize]{subfig}
\fi
\usepackage{url}
\usepackage{balance}
\usepackage{mathtools}
\usepackage{commath}
\usepackage{amssymb,amsbsy}
\usepackage{amsthm}
\newcommand{\Adb}{-90}

\begin{document}
	
	\title{RadioHound: A Pervasive Sensing Network for Sub-6~GHz Dynamic Spectrum Monitoring}

	\author{\IEEEauthorblockN{Nikolaus Kleber$^\dagger$, Jonathan Chisum$^\dagger$, Aaron Striegel$^\ast$, Bertrand Hochwald$^\dagger$,\\Abbas Termos$^\dagger$, J. Nicholas Laneman$^\dagger$, Zuohui Fu$^\ast$ and John Merritt$^\dagger$}
	\IEEEauthorblockA{$^\dagger$Department of Electrical Engineering\\
	$^\ast$Department of Computer Science and Engineering\\
	University of Notre Dame,
	Notre Dame, IN 46556\\ Email: \{nkleber, jchisum, striegel, bhochwald, atermos,  jnl, zfu1, jmerrit1\}@nd.edu}
	\thanks{This work was supported in part by the National Science Foundation through grant CNS-1439682 and by the Lab for Telecommunications Sciences.}%
	}
	
	\maketitle
	
	\begin{abstract}
		We design a custom spectrum sensing network, called RadioHound, capable of tuning from 25 MHz to 6 GHz, which covers nearly all widely-deployed wireless activity.  We describe the system hardware and network infrastructure in detail with a view towards driving the cost, size, and power usage of the sensors as low as possible.  The system estimates the spatial variation of radio-frequency power from an unknown random number of sources.  System performance is measured by computing the mean square error against a simulated radio-frequency environment.  We find that the system performance depends heavily on the deployment density of the sensors.  Consequently, we derive an expression for the sensor density as a function of environmental characteristics and confidence in measurement quality.
		
	\end{abstract}
	
	\section{Introduction}
	\subsection{Motivation}
	Knowledge of radio-frequency (RF) power in time, space, and frequency is valuable with several potential applications.  For example, such knowledge can inform spectrum policies and verify claimed spectral utilization.  Additionally, it can enable opportunistic spectrum access (OSA) through which a secondary user (SU) accesses the spectrum while the licensed primary user (PU) is not active on the band \cite{Zhao_Sadler_2007,Huang_et_al_2008}, which helps to improve spectrum utilization \cite{PCAST,NITRD,BerrySpectrumMarkets}.  Finally, knowledge of RF power can enable RSSI (radio signal-strength indicator) localization of transmitters \cite{Cheng_et_al_2012}, which itself has several potential applications, including the enforcement of spectrum policies.
	
\begin{figure}[bt]
		\centering
		\includegraphics[width=0.40\textwidth]{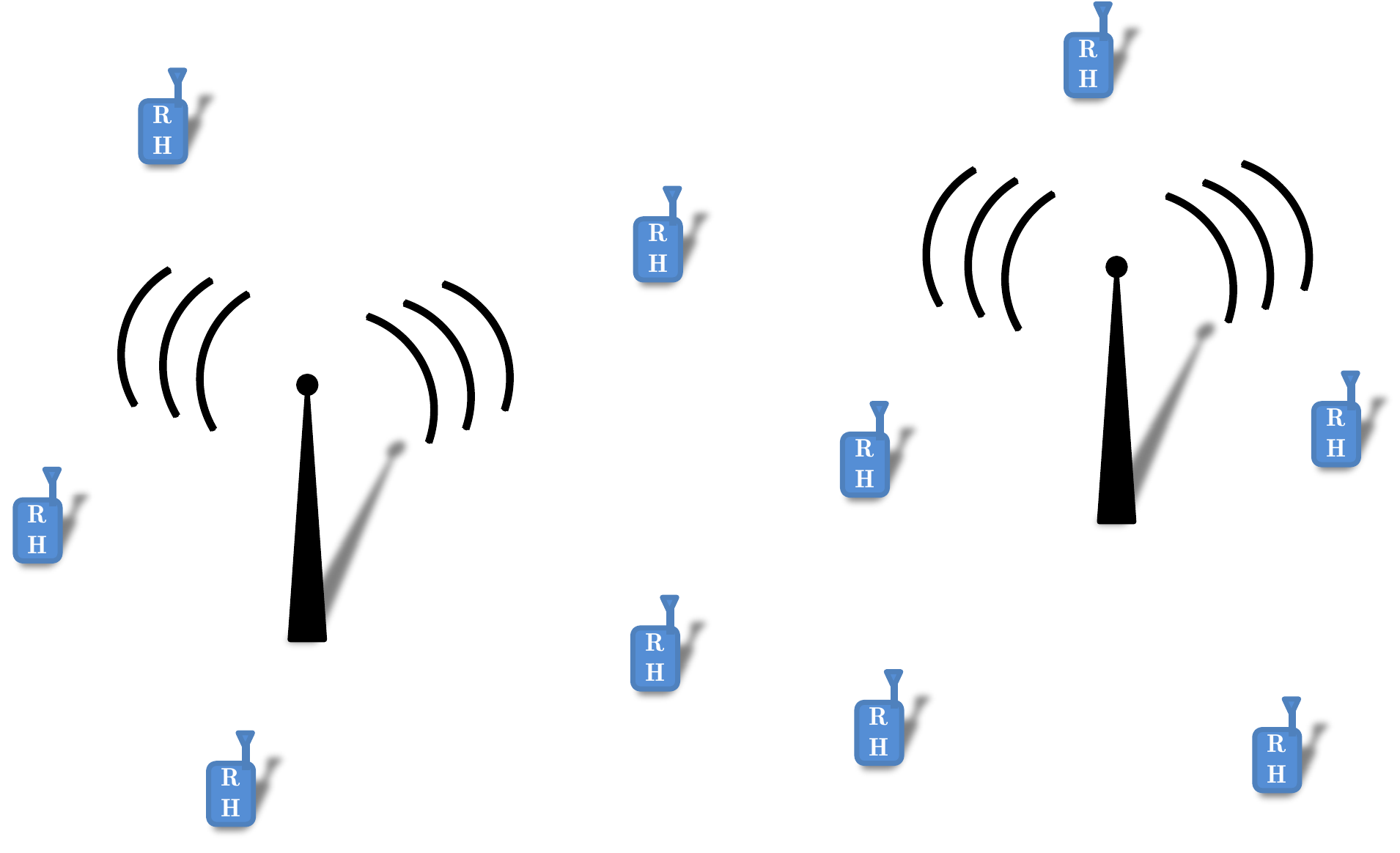}
		\caption{A network of sensors detecting RF power.}
		\label{fig:IntroConcept}
	\end{figure}

	To obtain information about RF power over time, frequency and space, a sensor network such as shown in Figure~\ref{fig:IntroConcept} is needed.  Wide-scale deployment with a large number of sensors is of particular importance for obtaining an accurate representation of activity.  For a given cost budget, it may be preferable to deploy a large number of low-performance sensors versus few high-performance sensors; however, quantification of this tradeoff needs study.  After describing the RadioHound system, we focus on the study of required sensor density.
	
	The RadioHound system can scan from 25 MHz to 6 GHz, the frequency range which contains the bands of interest for OSA.  The sensor network ``sniffs'' the spectrum for activity and records RF power usage in the form of periodograms tagged with a location and time, all of which is sent to a centralized database for further analysis and application.
	
	In an effort to drive down cost, size, and power consumption of the sensors, we need to assess what aspects of the network are important to maintain quality of the estimated spectral activity.  In this paper we focus on sensor density as one aspect that determines cost, and formulate a method to quantify sensor density requirements as a function of environmental characteristics and confidence in measurement quality.
	
	\subsection{Previous Work}
	
	Spectrum sensors that cover the desired sub-6~GHz band are commercially available.  The less expensive examples include the AD9364 RFIC and HackRF \cite{AD9364,HackRF}.  However, the cost of these sensors is still approximately \$200 to \$300.  Our near-term target is sub-\$40, with a long-term goal sub-\$10.  Our current sensors leverage low-cost, off-the-shelf RTL-SDRs with a custom printed circuit board (PCB) extension hosted by a Raspberry Pi (RP).  In order for a spectrum sensor network to be densely deployed, cost should not be a significant obstacle.
	
	The use of a RP in conjunction with an RTL-SDR as a low-cost spectrum sensor has been previously applied.  Gr\"{o}nroos et al. found comparable performance between a network of four such sensors and an expensive, professional RF sensor \cite{Gronroos_et_al_2016}. Additionally, IBM is utilizing the RP and RTL-SDR combination as part of their Internet-of-Things (IoT) project called Blue Horizons \cite{IBM_BlueHorizons}, in which a user can  generate waterfalls of spectrum activity and listen to an FM radio station in the locality of the sensor. The authors in \cite{nika2016empirical} use an RTL-SDR with a smartphone as a platform to investigate the feasibility of temporally monitoring the spectrum and localizing a transmitter based on \cite{yedavalli2005ecolocation}. Finally, the vendor Adafruit has instructions for developing a personal spectrum scanner from an RTL-SDR and RP \cite{DiCola_2015}.
	
	Our work differs in the following respects.  First, we have a custom-made front end that attaches to any off-the-shelf low-cost down-converter and digitizer such as the RTL-SDR.  We thereby achieve tuning range continuously to 6~GHz.  Second, we collect tagged time and location-dependent periodograms, and store them in a centralized server for later retrieval and synthesis.  In \cite{Gronroos_et_al_2016}, the authors focus on a the problem of detecting the presence of a transmission.  We focus instead on accurate reproduction of spectral activity, independently of the random number of sources present.
	
	\subsection{Contributions}
	
	The RadioHound system is described in detail in Section~\ref{sec:RadioHound System}.  We outline the components of the low-cost sensors and discuss the underlying infrastructure to connect the sensors, save data, and interact with users.  The deployment of the RadioHound system brings with it several questions on performance.  This paper explores the initial results on two related problems.
	
	First, to generate an RF power map, Section~\ref{sec:System Performance} considers two-dimensional interpolation of irregularly-spaced data points.  Although the general problem has been explored in, for example, \cite{Crain_1970,McLain_1976,Maude_1973,Shepard_1968}, we must decide how at a given location to condense or fuse the reported periodograms over time and frequency into a single value that bears statistical significance.  With this value, we can then apply any of the various methods of interpolation.  As we will see, deployment density of the sensors has a significant effect on this interporlation.
	
	Second, Section~\ref{sec:Deployment Density} examines the proper deployment density of sensors.  We define constraints based on desired qualities of the RF power map and derive an equation for the sensor density using a propagation model and tools from stochastic geometry.	
	
	\section{RadioHound System} \label{sec:RadioHound System}
	
\begin{figure}[bt]
		\centering
		\includegraphics[width=0.4\textwidth]{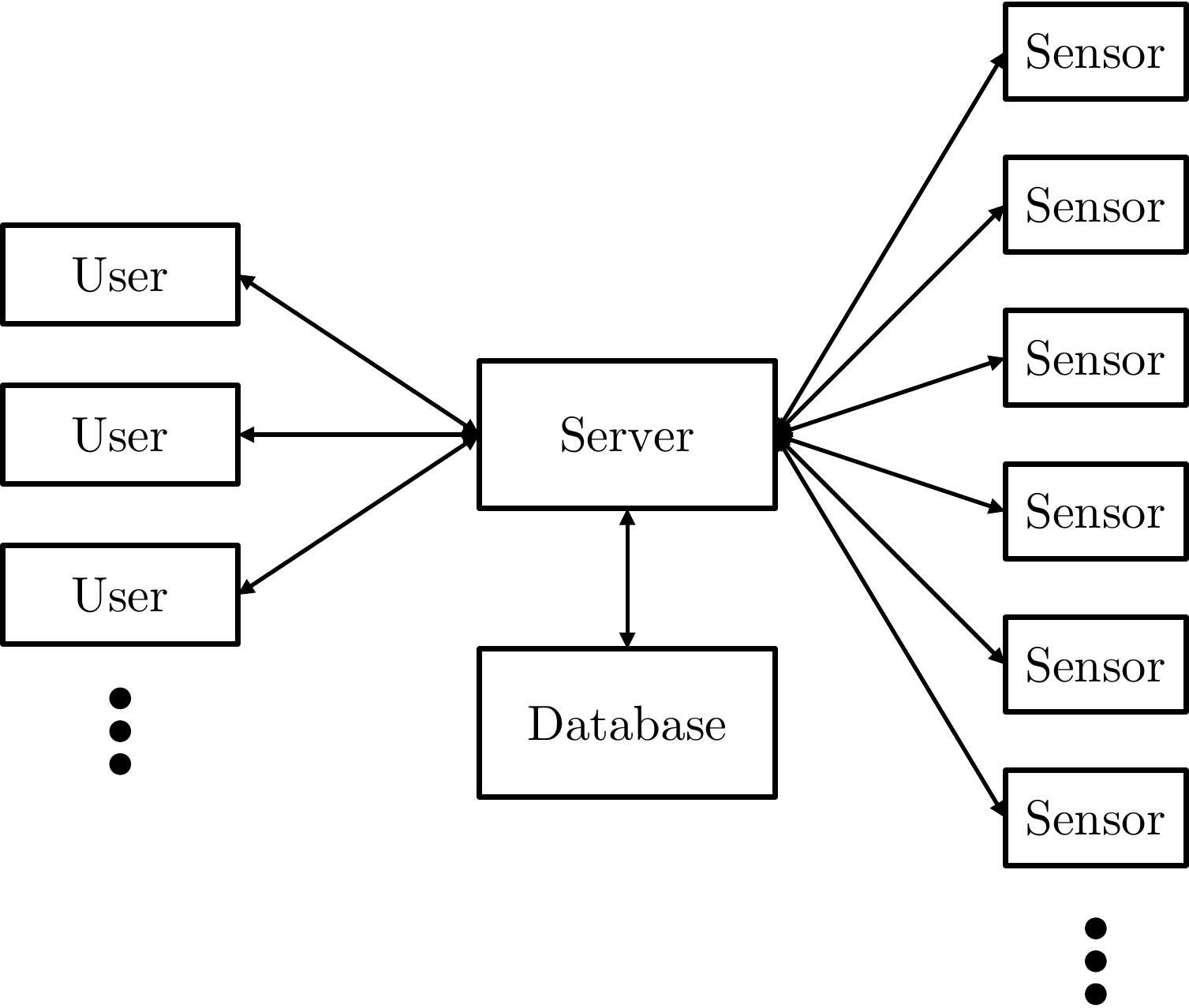}
		\caption{The interactions among the RadioHound system components.}
		\label{fig:ComponentInteractions}
	\end{figure}

	The RadioHound (RH) system contains three major components: a set of \textit{client sensors}, a \textit{centralized database/server/controller}, and a \textit{user interface}.  The high level interaction among these three components is shown in Figure~\ref{fig:ComponentInteractions}.  The user interface requests data for visualization or sends commands to the RH sensors via a server.  The client sensors receive commands from or report data to a server.  Consequently, the server is essentially a broker managing messages.  The server either stores data to or queries data from a database, or the server passes messages between the relevant parties.
	
	We present more details in the following sections for the sensor hardware, the software infrastructure, and the user interface of the RH system.
	
	\subsection{Sensor Design}
	
	The sensors consist, in part, of an RTL-SDR  hosted by a RP.  The RTL-SDR can scan from 25 MHz to 1.750 GHz \cite{Osmocom}, but we have created a custom front-end circuit that extends the range of the RTL-SDR to 6 GHz.  The RP host is a surrogate for a smartphone as the eventual host, and the RTL-SDR is a surrogate for a simple downconverter and digitizer that will eventually supplant.
		
\begin{figure}[bt]
		\centering
		\includegraphics[width=0.4\textwidth]{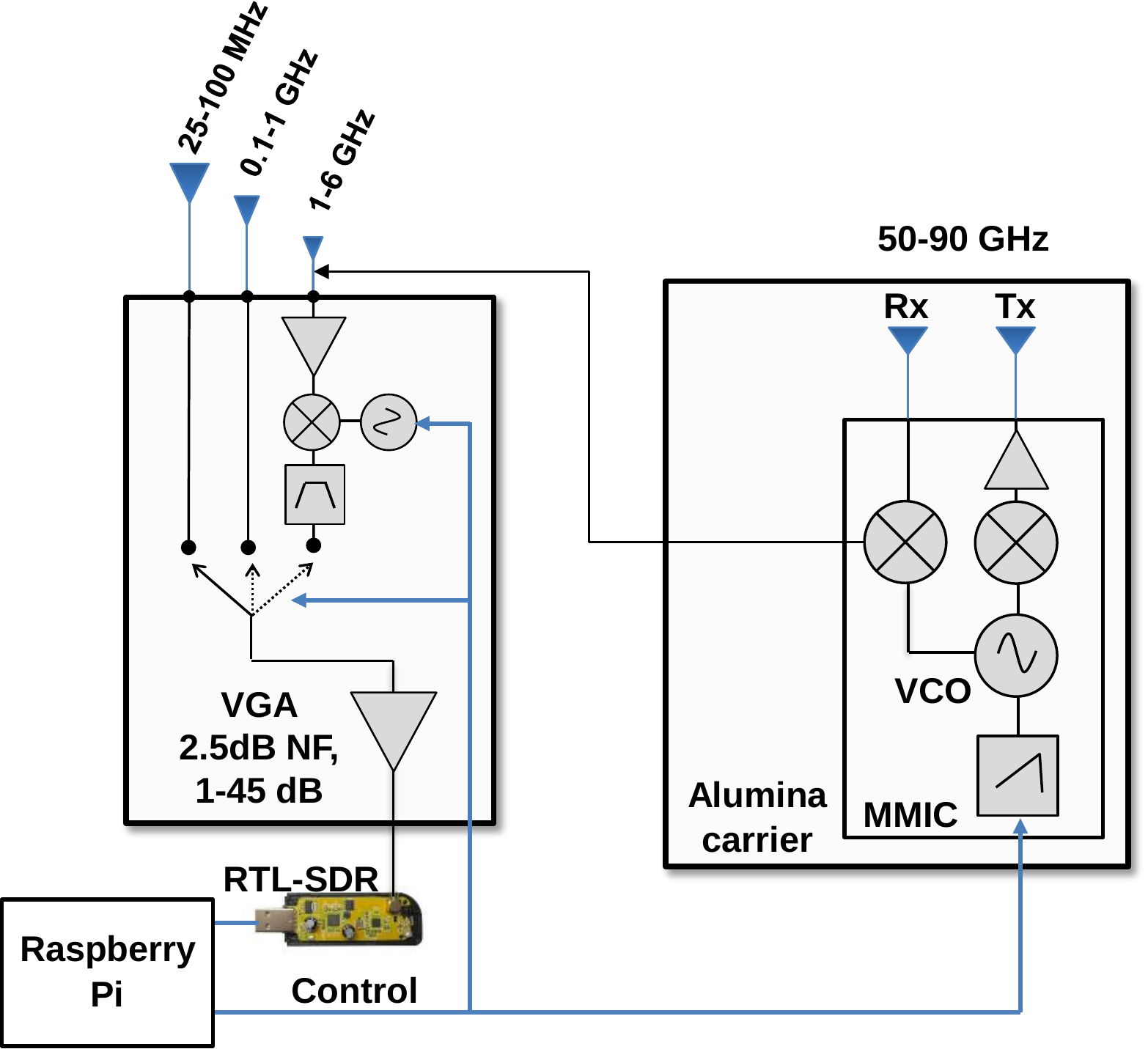}
		\caption{The RadioHound sensor hardware components.}
		\label{fig:RHsensorBlockDiagram}
	\end{figure}

	Figure~\ref{fig:RHsensorBlockDiagram} shows a block diagram of the RadioHound sensor hardware. The RP provides the interface to the controller, conducts sampling and on-node processing, and pushes samples to the database for visualization and further processing. The RTL-SDR acts as a fixed IF analog-to-digital converter and is capable of capturing 2 MHz of bandwidth with 8-bits of resolution. The custom RF front-end provides the tunable radio interface using a Hartley image-reject down-conversion stage to the IF. To support a variety of low-cost downconverters-digitizers (of which the RTL-SDR is an example) we selected a fixed IF of 110 MHz. The RF front-end provides wideband and high-speed tuning, down-conversion to the fixed IF, and variable amplification for increased dynamic range. In particular the custom RF front-end: 
	
	\begin{enumerate}
		\item Expands the tunable range to cover the 25 MHz to 6 GHz band and adds provisions for future millimeter-wave front-end modules covering the anticipated 5G bands up to 90 GHz.
		\item Decreases the tune/settling time of the receiver by adding a phase locked loop (PLL) and voltage controlled oscillator (VCO) with a settling time of $50\,\mu$s (\textit{e.g.}, the 25 MHz to 6 GHz band can be scanned in as little as $0.15\,$seconds).
		\item Expands the dynamic range of the receiver by including a low-noise amplifier and variable gain amplifier with approximately $32\,$dB of tunable gain.
	\end{enumerate}
	
	The RF band is split into three sub-bands: $25-100\,$MHz, $100-400\,$MHz and $0.4-6.0\,$GHz. This allows the sensor to exploit various antenna technologies to cover such a wide band without resorting to very large antennas. For example, the $25-100\,$MHz band can use a ferrite-loaded coil antenna, and the $0.4-6\,$GHz band might employ printed spiral antennas. 
	
	\begin{figure}[bt]
		\centering
		\includegraphics[width=0.45\textwidth]{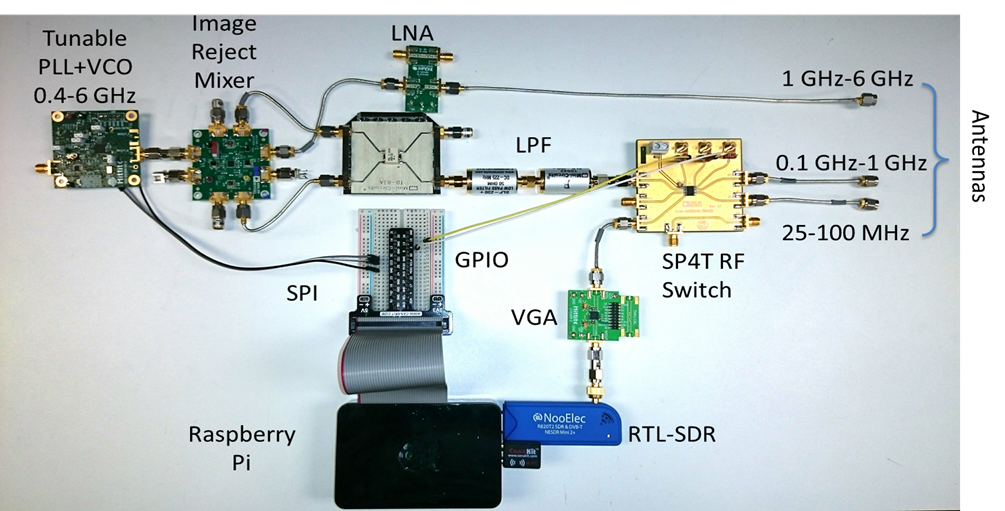}
		\caption{The prototype RadioHound sensor uses commercially available RFIC evaluation boards. The first deployed nodes will be integrated on a single printed circuit board.}
		\label{fig:RHsensorPhoto}
	\end{figure}
	
	Figure~\ref{fig:RHsensorPhoto} shows the prototype RadioHound sensor based on commercially available RFIC evaluation boards. The prototype is currently being integrated onto a single PCB with overall dimensions of $65\,\rm{cm}\times56.5\,\rm{cm}$, matching the RP and mounting holes placed for direct installation on top of the RP board.
	
	The final design of the sensor will incorporate the entire RF chain on the PCB board, eliminating the need for the RTL-SDR.  The final cost of a sensor will be sub \$10 and will have the flexibility to run from a variety of hosts, including laptops, RPs, and smartphones.
	
	The sensor in its current version consumes over 3 Watts.  The next version reduces the power consumption by half.  In order to maintain 24-hour battery life with a smartphone host, the final version should consume less than 0.5 Watts.
	
	We use low-cost components to implement the front-end. This results in many imperfections affecting the received signal such as LO drift, amplifier non-linearities, and IQ imbalance which reduces rejection of image signals. However, we postulate that such hardware imperfections will not significantly limit our system as we are interested in power measurements and not the demodulation of the data. A system with many scattered sensors allows fusion and compensation of the impaired measurements of any individual sensor. This idea of crowd sourcing RF measurements from a low-cost front-end is not new \cite{nika2014commoditized}, but further investigation is needed to quantify the performance of such a system.
	
	\subsection{Software Infrastructure}
	
\begin{figure}[bt]
		\centering
		\includegraphics[width=0.4\textwidth]{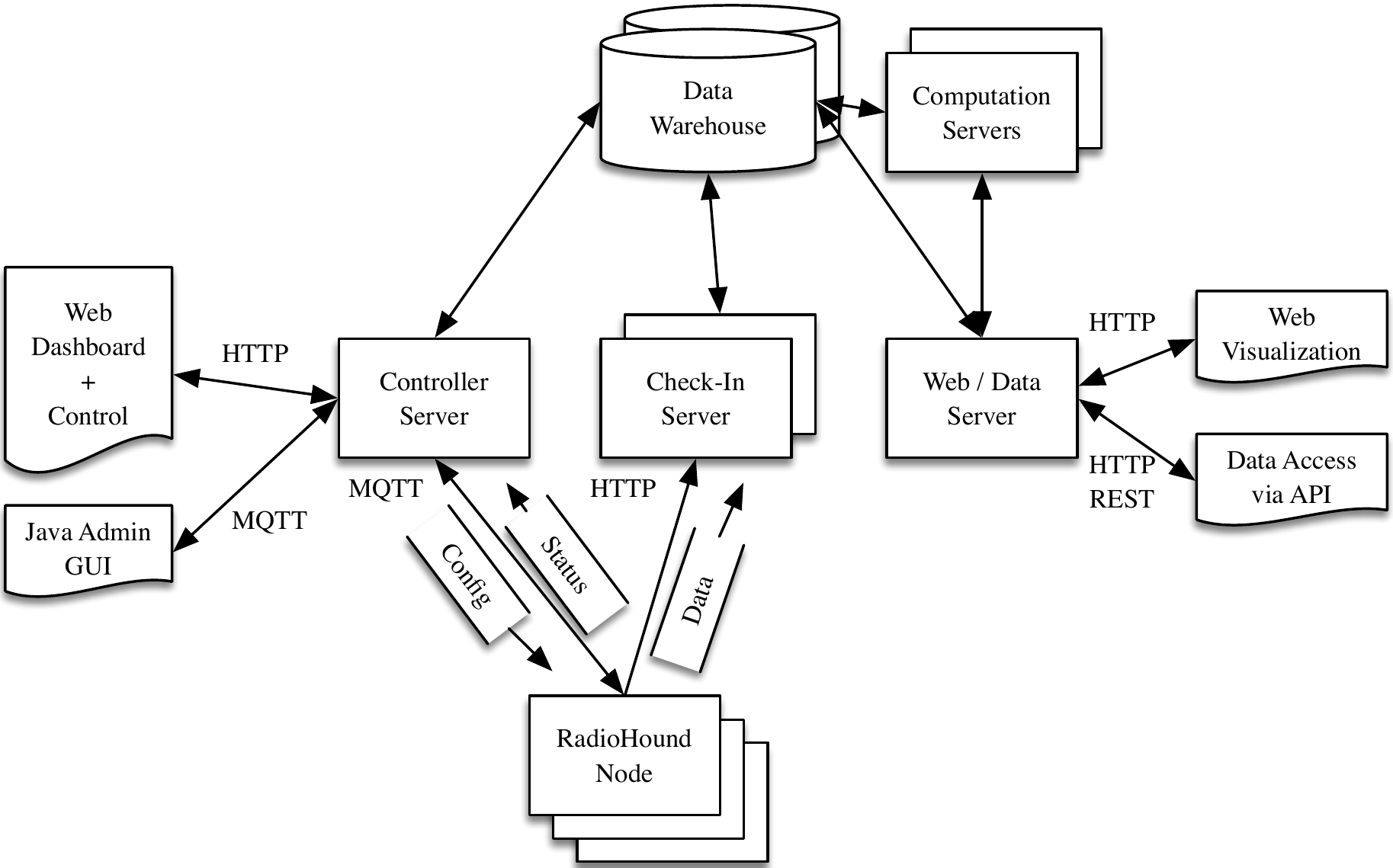}
		\caption{Overall RadioHound software infrastructure.}
		\label{fig:RH-SW}
	\end{figure}

\begin{figure}[bt]
		\centering
		\includegraphics[width=0.4\textwidth]{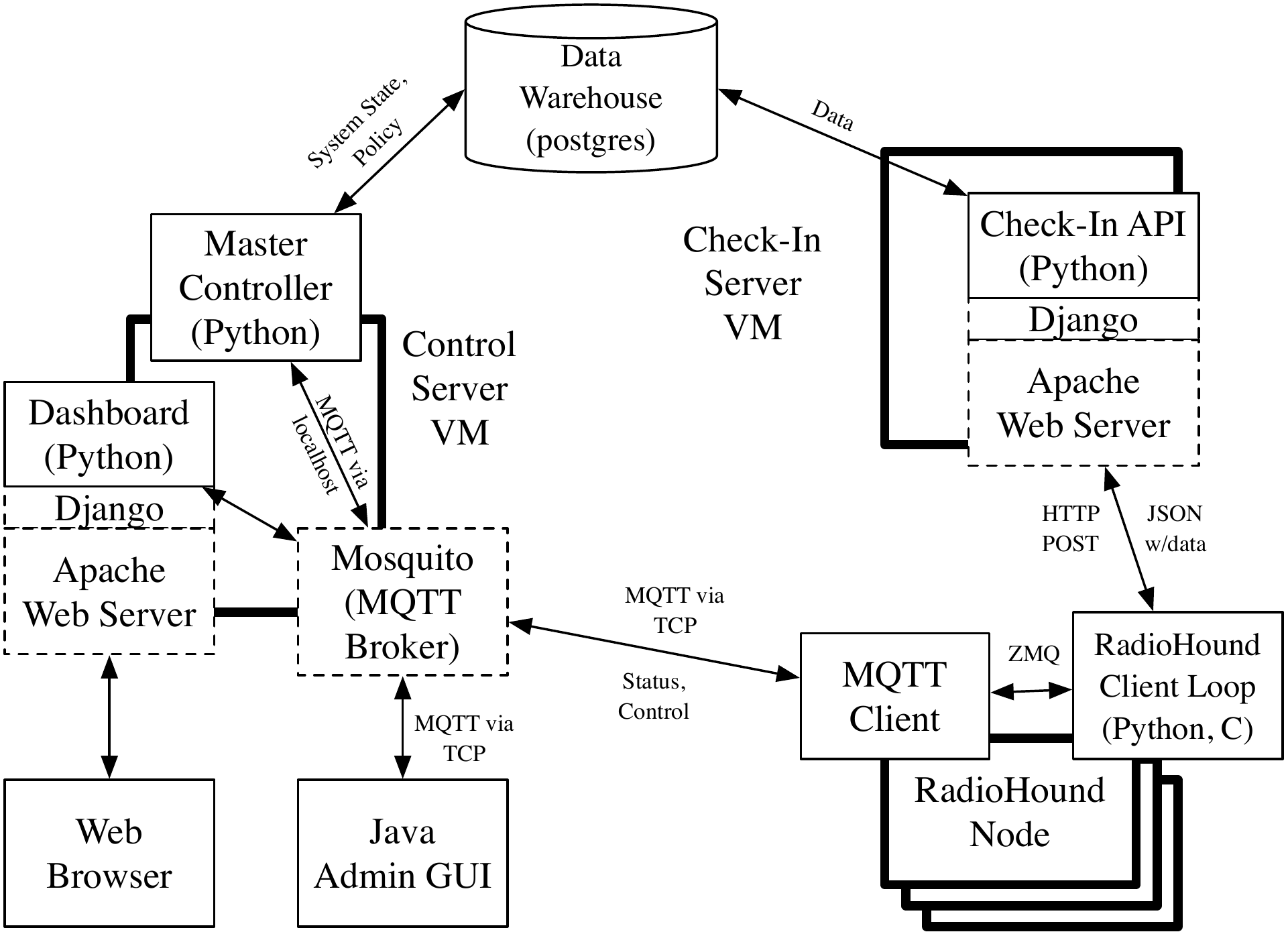}
		\caption{Control and data interactions for RadioHound nodes.}
		\label{fig:RH-MQTT}
	\end{figure}

	Figure~\ref{fig:RH-SW} presents a high level overview of the RadioHound software infrastructure which can be subdivided into three primary components: control (management), sensing (RadioHound nodes), and the data back-end (storage, querying, visualization). Figure~\ref{fig:RH-MQTT} breaks down the control and sensing aspects of RadioHound in further detail.  
	
	Each RadioHound node runs a local software loop responsible for processing jobs as given by the Master Controller.  A job is nominally a request for scanning across a particular frequency range along with the requisite settings for conducting the scan as well as intermediate processing.  The local software loop takes commands via an MQTT client that attempts to maintain a persistent TCP connection with the MQTT server running at the Control server. MQTT is used because it is a simple publish / subscribe protocol with human readable strings and numerous open-source implementations that is frequently used for IoT systems .  
	
	A localized broker running at the client accepts the MQTT commands and distributes the commands to a localized processing loop of C code and Python code using ZMQ (Zero Message Queueing) as a connector.  Native C code is used for interacting with the RTL-SDR device (and accompanying RadioHound hardware).  Python processing is used for conversion of the raw data into a periodogram via the Welch method \cite{Proakis}.  The periodogram is then conveyed via an HTTP POST to the Check-In server as a JSON.
	
	The Check-In server (realized via Django) is responsible for maintaining a connection to the Data Warehouse (storage) for storing the information in the Postgres database.  Postgres was selected for scaling purposes due to the potential for extremely large numbers of data records.  The Data Warehouse also serves as a data visualization front-end which provides data both through a web interface as well as data directly for download via an HTTP REST interface.  A stand-alone GUI for visualization has also been developed that leverages the REST interface.     
	
	\subsection{User Interface}
	
	The user interface acts as a means of controlling sensors and various pieces of the RadioHound system functionality. Additionally, the user interface acts as the means by which a person can visualize the collected data.  Given that RF power varies in time, frequency, and space, three key visualizations of the data are of interest.  First, a waterfall diagram demonstrates the variability of RF power over time for a fixed frequency range and location.  Second, a periodogram demonstrates variation in frequency for a fixed time and location.  Finally, an RF power map enables a user to see variation over space for a fixed frequency and time range.  All three visualizations will be available from the user interface, but our focus for this paper will be on the RF power map.   
	
	\section{RF Power Map} \label{sec:System Performance}
	
	One capability of the RadioHound system is estimating RF power as a function of space based on sensor measurements.  A user can clearly visualize the spatial variability of spectrum utilization via an RF power map.  Consequently, we evaluate the system performance by generating RF power maps through a simulated RF environment in which the true RF power is known.  We then determine the mean square error (MSE) of the RF power maps measured across a rectangular mesh grid.
	
	\subsection{Current Method}
	
	First, we discuss our current method of estimating the RF power as a function of location, which we achieve through interpolation.  In general, we need triples $\left(x,y,z\right)$ in order to generate a surface.  The $x$ and $y$ coordinates are the longitude and latitude, respectively.  The $z$ value needs to be a statistically significant value relating to the power measured at that location over the specified frequency and time range.  Currently, we integrate each of the collected periodograms over the specified frequency range and compute the average value across periodograms.  We discuss this issue more in Section~\ref{subsec:Future Work System Performance}.
	
	As we mentioned previously, numerous methods of interpolation of irregularly-spaced data exist.  Currently, we implement a simple planar interpolation that we briefly describe here.
	
\begin{figure}[bt]
		\centering
		\includegraphics[width=0.45\textwidth]{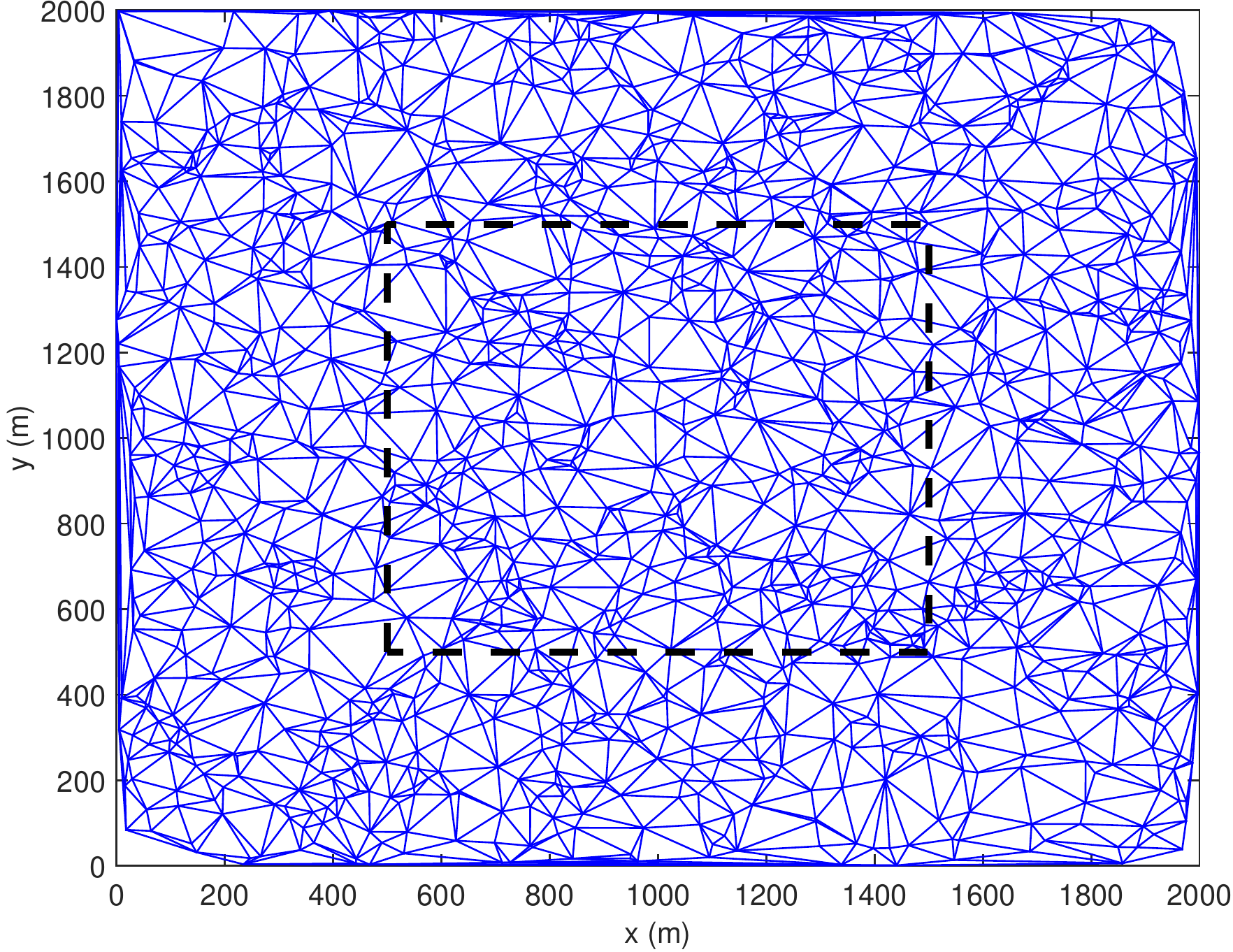}
		\caption{Delaunay triangulation of the $x$-$y$ plane.  The black outline shows the region taken for the RF power maps in Figure~\ref{fig:RF Power Maps}.} 
		\label{fig:DelaunayTriangulation}
	\end{figure}

	First, a two-dimensional Delaunay triangulation is performed on the $(x,y)$-coordinates of the data \cite{Lee_Schachter_1980}.  See Figure~\ref{fig:DelaunayTriangulation} as an example that we will expand upon.  The result is a partitioning of the plane into triangular regions with data points as the vertices of each triangle.  For each triangular partition, we calculate the planar equation $f(x,y) = ax +by + c$ that passes through the data points at the vertices.  Consequently, we have a piecewise continuous function $f(x,y)$ from which we can generate interpolated values.  Note that we perform planar interpolation on the dBm values of $z$ in order to better approximate the polynomial nature of path loss.
	
	\subsection{RF Environment Simulation}
	
	Now we describe the simulation of the RF environment, which is based in part on stochastic geometry.  We present more details of stochastic geometry in Section~\ref{sec:Deployment Density}, but here it suffices to say that a homogeneous Poisson point process (PPP) for a specified region has a Poisson random number of points that are independently, uniformly distributed over the region \cite{Haenggi_Stog}.  
	
	First, we choose a 1 km$^2$ square region and simulate the location of the sources as a two-dimensional, homogeneous PPP with intensity $\lambda_T = 3$ sources/km$^2$, which means we  have an average of three sources per square km.  We assume that each source transmits isotropically with a power of 30~dBm on the same frequency.
	
	To simulate the value of the RF power everywhere, we use the path loss law $\ell(r) = Kr^{-\alpha}$ with $\alpha=3$ and $K$ set according to the example in Section \ref{subsec:Density Example}.  We do not simulate shadowing or fading here.
	
	Next, we simulate the locations of the sensors as another homogeneous PPP with intensity $\lambda_S$.  For each point, we use the path loss model to calculate the sum power from the sources that is measured by the sensor.  From these measurements, we interpolate a fine mesh grid of values using the method described above.
	
\begin{figure}[!hbt]
		\centering
		\subfloat[Sensor density $\lambda_S=94$ sensors/km$^2$ (or $\beta = 0.5$ in \eqref{eq:NoLambdaT}).]{\centering \includegraphics[trim=1.5cm .5cm 1.5cm .5cm, clip = true, width=0.43\textwidth]{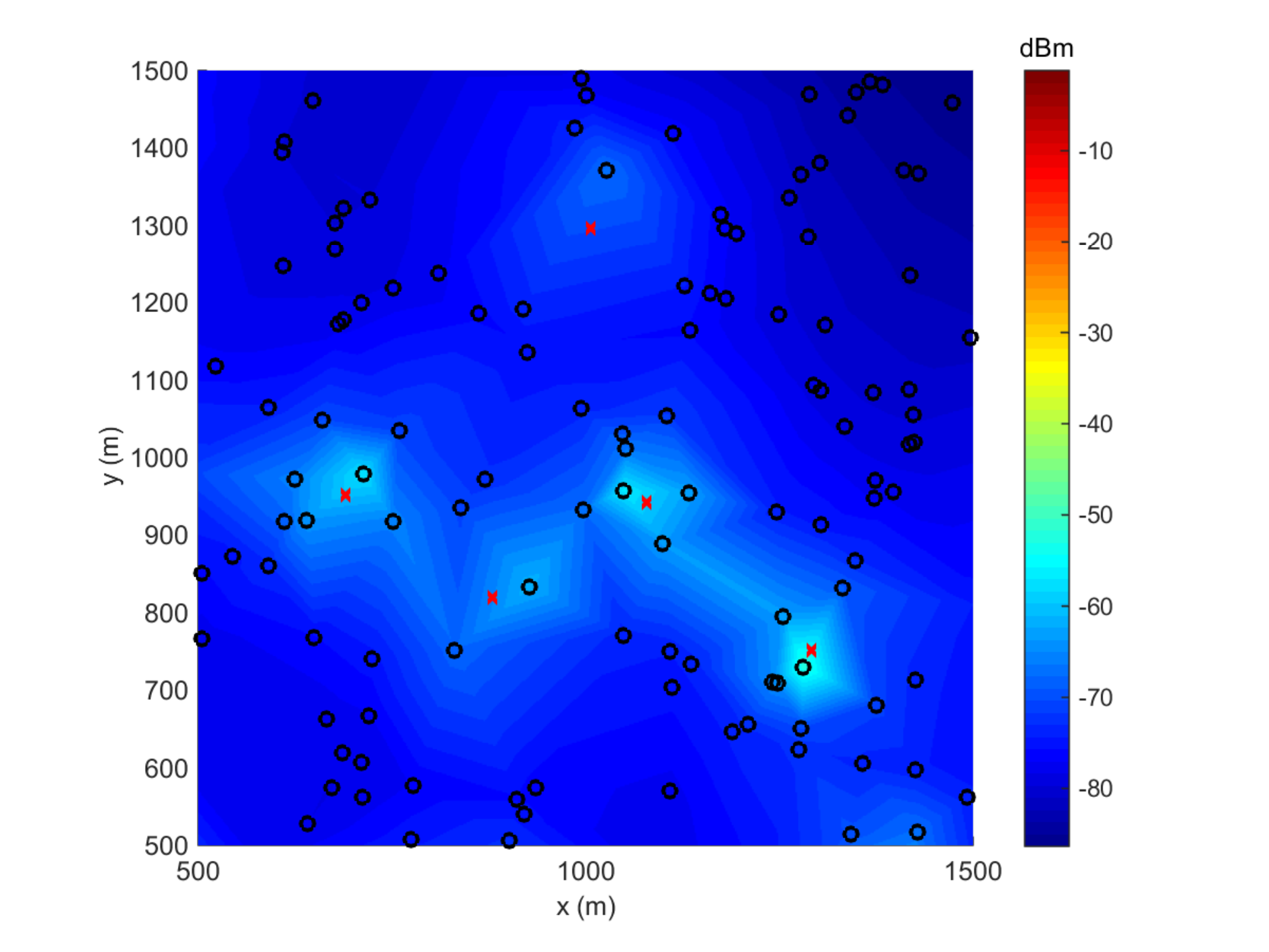} \label{fig:LinearInterpolationSamller}}
		
		\subfloat[Sensor density $\lambda_S=313$ sensors/km$^2$ (or $\beta = 0.9$ in \eqref{eq:NoLambdaT}).]{\centering	\includegraphics[trim=1.5cm .5cm 1.5cm .5cm, clip = true, width=0.43\textwidth]{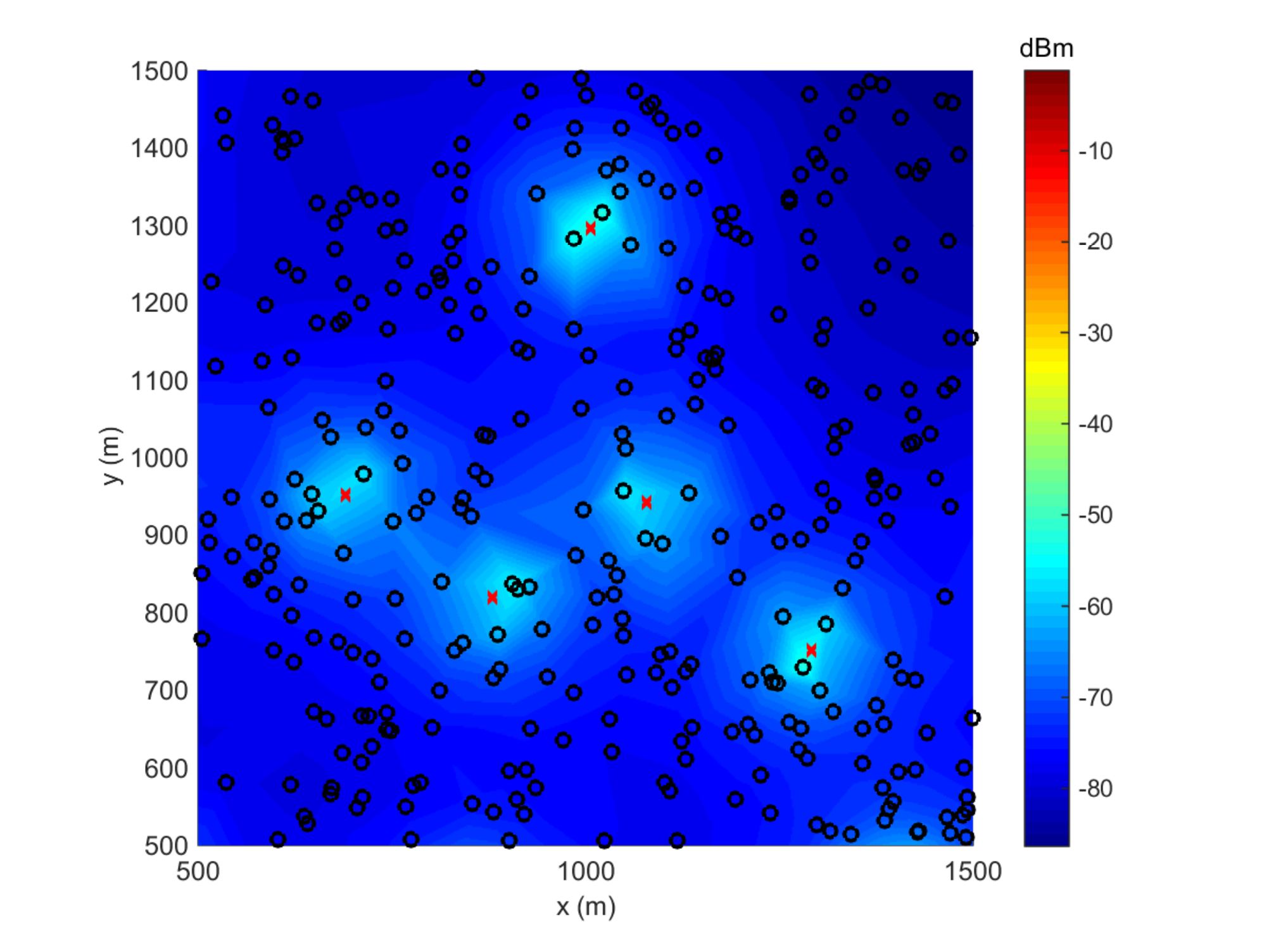} 	\label{fig:LinearInterpolation}}
		
		\subfloat[True RF power map.]{\centering \includegraphics[trim=1.5cm .5cm 1.5cm .5cm, clip = true, width=0.43\textwidth]{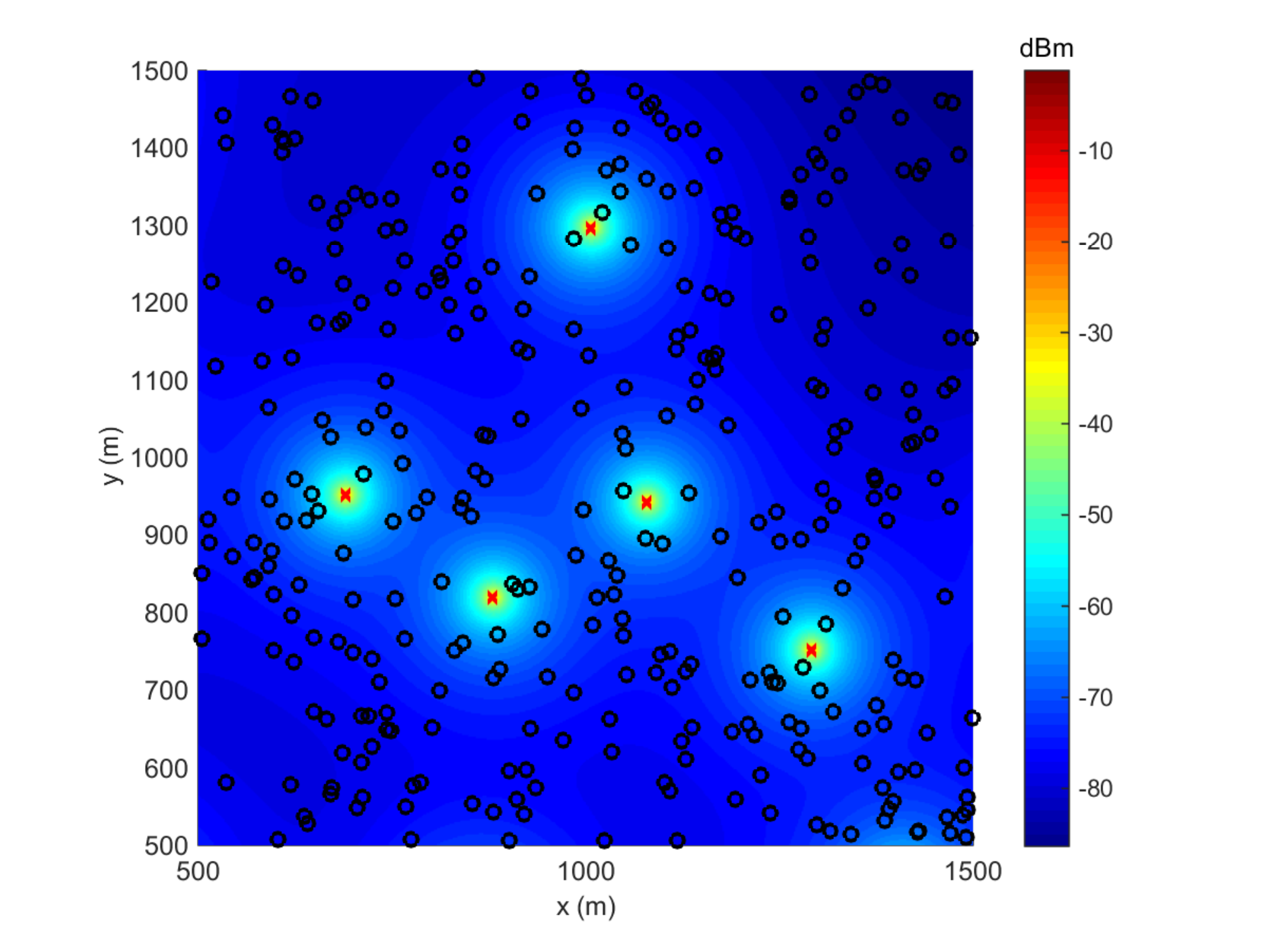} 	\label{fig:TrueHeatMap}}
		\caption{The RF power maps for (a) a sensor density of 94 sensors/km$^2$ (a 50\% confidence of detecting sources), (b) a sensor density of 313 sensors/km$^2$ (a 90\% confidence of detecting sources), and (c) the full simulated values.  A red `x' is the location of a source, and a black `o' is the location of a sensor.}
		\label{fig:RF Power Maps}
	\end{figure}

	We look at two cases: $\lambda_S = 94$ and $\lambda_S = 313$ sensors/km$^2$.  (These values correspond to a 50\% and 90\% confidence, respectively, that we will calculate in Section~\ref{sec:Deployment Density}.)  The results are in Figure~\ref{fig:RF Power Maps}, and the true RF power map is given in Figure~\ref{fig:TrueHeatMap}.
	
	The measured MSE between the results in Figures~\ref{fig:LinearInterpolationSamller} and \ref{fig:TrueHeatMap} is 6.99, while the MSE between the results in Figures~\ref{fig:LinearInterpolation} and \ref{fig:TrueHeatMap} is 2.40.  Thus, increasing the sensor density by approximately a factor of 3 reduced the MSE by approximately a factor of 3.
	
	\section{Deployment Density} \label{sec:Deployment Density}
	
	From the previous section, we see that sensor density greatly impacts the quality of the RF power interpolation across space.  Consequently, we would like to explicitly relate sensor density with a confidence in the quality of the RF power map.
	
	In general, the problem of determining a proper sensor density for a ``good'' RF power map is ill-posed.  As a result, we place a spatial constraint on problem informed by our current interpolation method.  Since the interpolation values can never be larger than the maximum value of the sampled data, we need at least one sensor to be geographically close to each source.
	
	If a sensor is not close enough to a given source, the source is essentially invisible on the planar-interpolated RF power map.  On the other hand, one well-placed sensor can detect the power of two or more proximate sources for an RF power map, though a spatial resolution problem results.  In other words, if one sensor is between two nearby sources, the RF power map will appear to have one large source.  To ensure proper localization of sources without resolution problems requires a higher density of sensors.  If we limit our goal to finding opportunities for spatial reuse of the spectrum, we only need a general idea of the location of the RF source(s) without necessarily needing to resolve the location of each antenna.  This will allow for a lower density of sensors.  Consequently, for the sake of obtaining an estimate of sensor density, we argue for at least one sensor to be geographically close to each source, though one sensor can cover multiple sources.
	
	\subsection{Spatial Constraint}
	
	We enforce the constraint that a sensor should, with high probability, be within a radius $r$ of any source, or else the source will not be detected by any sensor.  We assume the source locations are unknown and random and therefore model the locations as a stochastic point process.  Similarly, since the deployment of sensors will be based on an uncoordinated crowd of volunteers, we cannot influence their locations, even though we assume we know them.  Consequently, we also model the sensor locations as a stochastic point process.  In particular, since the Poisson point process (PPP) models the independent random movement of mobile users and the deployment of base-stations fairly well \cite{Haenggi_Stog}, we will model the irregularly spaced sources and sensors as PPPs.  
	
	We anticipate that in reality, both sensors and sources will form cluster point processes since people tend to congregate in some desirable areas while avoiding other areas, such as lakes, canyons, or ``bad'' parts of town.  The results based on the PPP will offer us the density, on average, that we will need.
	
	We use the notation found in \cite{Haenggi_Stog}.  Let $\Phi_T, \ \Phi_S \subset \mathbb{R}^2$ be independent, homogeneous PPPs with intensities $\lambda_T$ and $\lambda_S$, respectively.  $\Phi_T$ and $\Phi_S$  represent the locations of the sources (transmitters) and sensors, respectively.
	
\begin{figure}[bt]
		\centering
		\subfloat[Disk around source.]{\centering \includegraphics[width=0.2\textwidth]{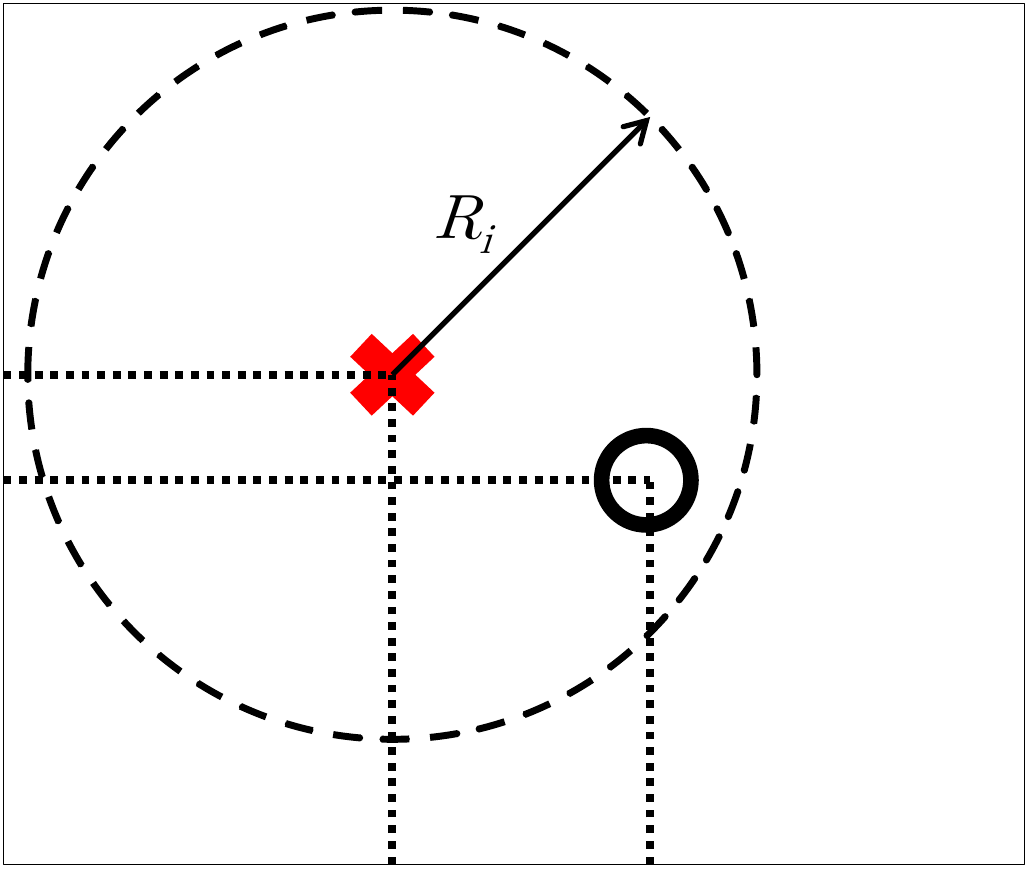} \label{fig:SwitchCirclesA}}
		\subfloat[Disk around sensor.]{\centering
			\includegraphics[width=0.2\textwidth]{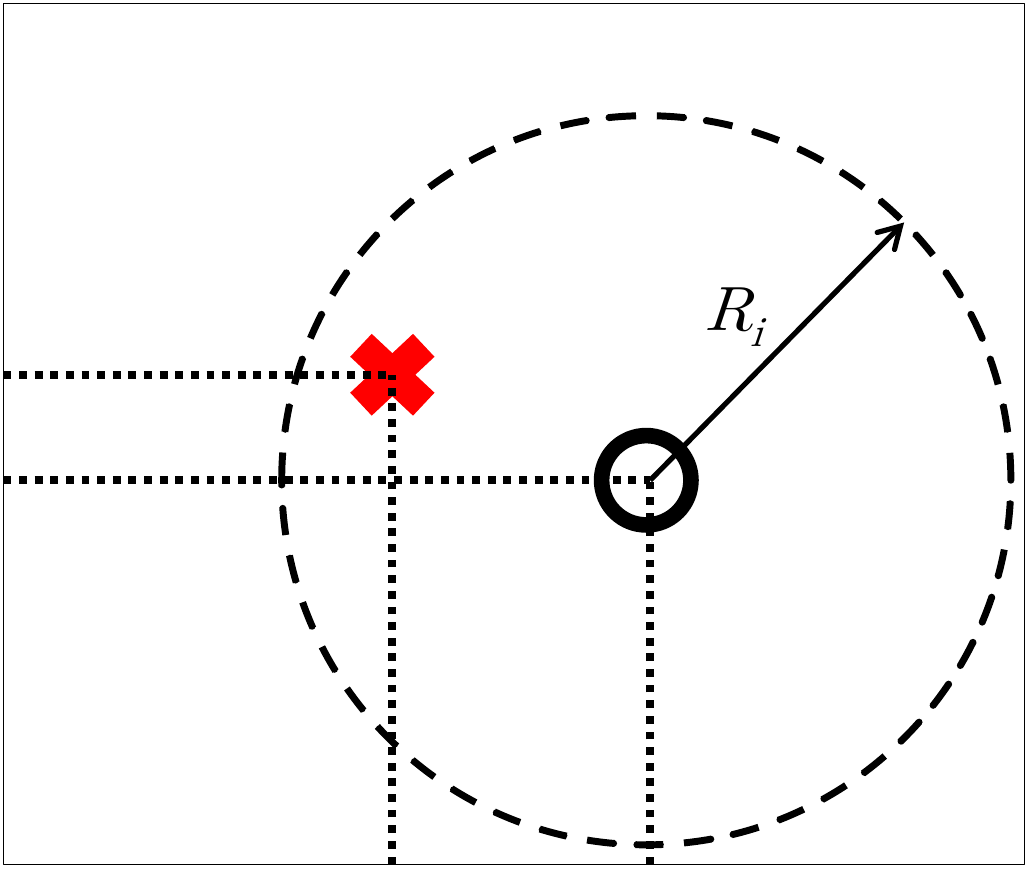} 	\label{fig:SwitchCirclesB}}
		\caption{The reversal of roles in the coverage problem.  The red `x' is the location of a source, and the black `o' is the location of a sensor.}
		\label{fig:Disks}
	\end{figure}

	Naturally, we would like to draw disks around the sources and determine the probability that at least one sensor falls within each disk.  However, it is instructive to reverse the roles of the sensors and sources to cast a coverage problem.  After all, if a sensor is within the disk of a source, that source also falls within a disk of the same radius around the sensor.  See Figure~\ref{fig:Disks}.  Now our desired result is to have all sources fall within the union of disks around the sensors.
	
	Mathematically, let
	\begin{align}
	\Xi \triangleq \bigcup_{i:y_i\in \Phi_S} b(y_i,r)
	\end{align}
	be the union of the disks with radius $r$ centered around the points of $\Phi_S$.  Since the regions (the grains) surrounding the point process are independent and identically distributed (IID) and the points in $\Phi$ (the germs) form a uniform PPP, $\Xi$ is a Boolean model \cite[Def.~13.4]{Haenggi_Stog}, which is a sub-class of germ-grain models.
	
	Next, we want to know the necessary density of sensors for a given probability that an arbitrary source will be covered by the sensors.  More explicitly, let $\mathcal{E}$ be the event that an arbitrary point in $\Phi_T$ is within the coverage region of the sensors: $\mathcal{E} = \{x \in \Phi_T: x \in \Xi\}$.  We want to find $\mathbb{P}(\mathcal{E}) = g(\lambda_S) > \beta$, where $g$ is some function, and solve for $\lambda_S$ in terms of $\beta$.
	
	Fortunately, there is a useful theorem regarding coverage and vacancy in Boolean models that will make our calculation relatively straightforward.  Specifically, for a Boolean model with a PPP of intensity $\lambda$ and IID grains $S$, the probability that a location is not covered is $\exp(-\lambda \mathbb{E}|S|)$ \cite[Th.~13.5]{Haenggi_Stog}.  This is also the fraction of a region not occupied by the union.
	
	Applying this theorem yields
	\begin{align}
	\mathbb{P}(\mathcal{E}) & = 1 -  \mathbb{P}(\mathcal{E}^C) \nonumber \\
	& = 1 - \exp\left(-\lambda_S \mathbb{E}|S|\right)\nonumber \\
	& = 1 - \exp\left(-\lambda_S \pi r^2\right) \label{eq:Spatial Constraint Probability}
	\end{align}
	Solving for $\lambda_S$ in the inequality $\mathbb{P}(\mathcal{E}) > \beta$ gives us
	\begin{align}
	\lambda_S > -\frac{\ln(1-\beta)}{\pi r^2}. \label{eq:lambdaS for spatial constraint}
	\end{align}
	
	The choice of $r$ remains.  We next translate the spatial requirement of $r$ into a relative power requirement.
	
	\subsection{Relative Power Constraint}
	
	The spatial constraint translates into a power constraint when an RF propagation model is considered.  For a sensor to have a specified receive power relative to the source power, the sensor must lie within some region near the source.  The size and shape of this region depends on the path loss, fading, and shadowing. 
	If we denote the sensor received power as $P_R$ and the source transmit power as $P_T$, we state this constraint as $(P_R/P_T)_{\text{dB}} \ge -A$~dB.
	
	Here we make the following simplifying assumptions about the propagation environment.  We will only consider path loss and shadowing while ignoring fading for ease of exposition and a first-order estimate.  Additionally, we assume that the sources and sensors have omni-directional antennas.  Finally, we assume that all sources transmit with high enough power such that -$A$ dB below the source power is above the receiver sensitivity of the sensor.   
	
	Next, let $\ell(r)$ be the path loss function, where $r$ is the distance between a sensor and a source, \textit{i.e.} $r = \lVert x_i - y_j\rVert$ where $x_i \in \Phi_T$ and $y_j \in \Phi_S$.  In general, many path loss functions exist \cite{Haenggi_Stog,Goldsmith_Wireless_Communications_2005}, and $\ell(r)$ can be changed as desired. We define $\ell(r) \triangleq K r^{-\alpha}$ for this exposition, where $K$ is a constant based on antenna characteristics and frequency, and $\alpha$ is the path loss exponent with a value usually between 2 and 4.
	
	To model the effects of shadowing, we let $Z_i$ be IID log-normal random variables with parameters $\mu$ and $\sigma$.  We can model the received power at sensor $i$ as $P_R = P_T\ell(r_i)/Z_i$.  In decibels, $P_{R,\text{dB}} = P_{T,\text{dB}} + 10\log_{10}K-10\alpha\log_{10}r-Z_{i,\text{dB}}$, where $Z_{i,\text{dB}}$ has a Gaussian distribution $\mathcal{N}(\mu,\sigma^2)$ \cite{Goldsmith_Wireless_Communications_2005}.
	
	With this function, we can determine a spatial requirement for the locations of the sensors relative to the sources, keeping in mind that the sources transmit isotropically.  In general, if we want a sensor within $-A$~dB of the peak power of a source, then we have $-A = 10\log_{10}\left(\frac{P_R}{P_T}\right) = 10\log_{10}\left(\ell(r_i)/Z_i\right)$.  Solving for $r_i$ for our particular path loss function gives us $R_i = \left(K10^{A/10}\right)^{1/\alpha}Z_i^{-1/\alpha}$.  Hence, a sensor must be within a random distance $R_i$ of the source.  If we consider this requirement in two-dimensional space, the sensor must consequently lie within some region $S$, which has a random shape since shadowing is direction dependent.
	
	From the theorem cited in the previous section, we will only be concerned with the average area of the $S$.  This will help us make some simplifications.  First, let us consider the two dimensional space in polar coordinates.  We know the radius is a function of the angle $\theta$ due to shadowing.  Let us divide the region into sectors by dividing $2\pi$ into $n$ equal sectors.  Next, let us determine the average radius $R_j$ for each sector $j$.  The area of $S$ becomes $|S| = \sum_{j=1}^{n}\frac{\pi R_j^2}{n}$.  To determine the average area of $S$, we have
	\begin{align}
	\mathbb{E}|S| & = \mathbb{E}\left(\sum_{j=1}^{n}\frac{\pi R_j^2}{n}\right) \nonumber \\
	& = \frac{\pi}{n} \sum_{j=1}^{n} \mathbb{E}\left(R_j^2\right) \nonumber \\
	& = \pi \mathbb{E}\left(R_j^2\right), \label{eq:Average of Sectorization}
	\end{align}
	where the last equality follows from $Z_j$ being IID.
	 
	The result in \eqref{eq:Average of Sectorization} shows that the average area does not depend on $n$.  As a result, we can simplify the problem to be a sensor falling within a random radius $R_i$ of the source.
	
	We will again use the germ-grain model, where now
	\begin{align}
	\Xi \triangleq \bigcup_{i:y_i\in \Phi_S} b(y_i,R_i)
	\end{align}
	is the union of the disks with IID random radius $R_i$ centered around the points of $\Phi_S$.  Again, the regions surrounding the point process are IID, so $\Xi$ is a Boolean model for which we can apply the previously mentioned theorem.
	
	For this case, from \eqref{eq:Average of Sectorization} we have
	\begin{align}
	\mathbb{E}|S| &= \pi \mathbb{E}( R^2) \nonumber \\
	&= \pi\mathbb{E}\left( \left((K10^{A/10})^{1/\alpha}Z^{-1/\alpha}\right)^2\right) \nonumber \\
	&= \pi K^{2/\alpha}10^{A/5\alpha} \mathbb{E}\left(Z^{-2/\alpha}\right) \nonumber \\
	& = \pi K^{2/\alpha} 10^{A/5\alpha}  \exp\left(-2\zeta \mu /\alpha +2\zeta^2 \sigma^2/\alpha^2\right), \label{eq:E|S|}
	\end{align}
	where $\zeta = \ln 10 / 10$ is a scaling factor due to the base 10 of the logarithm.
	
	Substituting \eqref{eq:E|S|} into \eqref{eq:Spatial Constraint Probability} results in
	\begin{align}
	\mathbb{P}(\mathcal{E}) & = 1 - \exp\left(-\lambda_S \pi K^{2/\alpha} 10^{A/5\alpha}  \text{e}^{\left(-2\zeta \mu  /\alpha +2 \zeta^2 \sigma^2/\alpha^2\right)}\right).
	\end{align}
	Again, solving for $\lambda_S$ in the inequality $\mathbb{P}(\mathcal{E}) > \beta$ yields
	\begin{align}
	\lambda_S > -\frac{\ln(1-\beta)}{\pi K^{2/\alpha} 10^{A/5\alpha}  \text{e}^{\left(-2\zeta \mu/\alpha +2\zeta^2 \sigma^2/\alpha^2\right)}}. \label{eq:NoLambdaT}
	\end{align}
	
	\subsection{Example} \label{subsec:Density Example}
	
\begin{figure}[bt]
		\centering
		\includegraphics[width=0.475\textwidth]{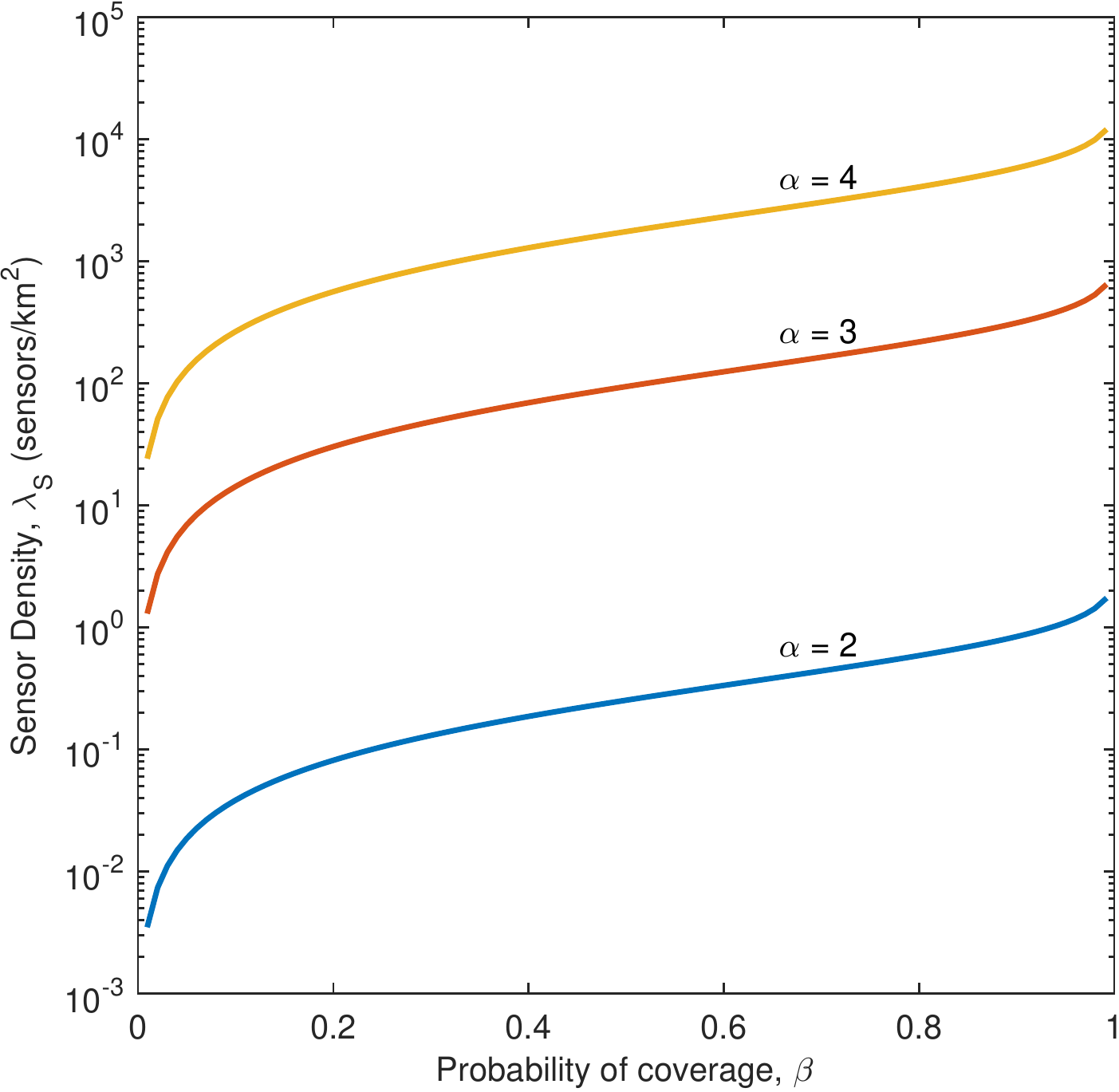}
		\caption{Sensor density as a function of the probability ($\beta$) that an arbitrary source is within radius $R_i$ of any sensor at a 1~GHz frequency.  Given by \eqref{eq:NoLambdaT}.}
		\label{fig:LSvsBeta_NoLambdaT}
	\end{figure}

\begin{figure}[bt]
		\centering
		\includegraphics[width=0.475\textwidth]{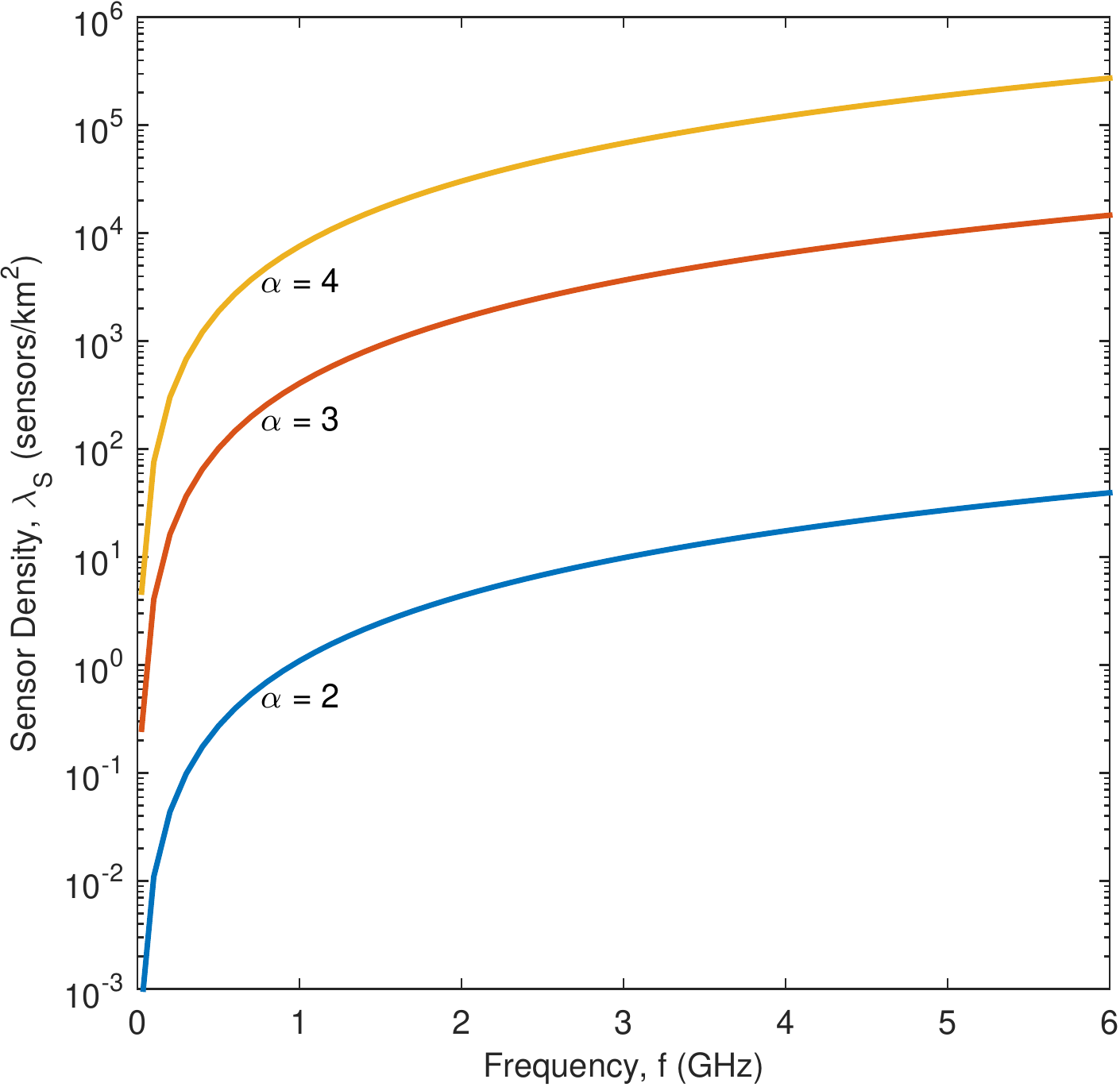}
		\caption{Sensor density as a function of the frequency  with $\beta=0.95$.  Given by \eqref{eq:NoLambdaT} with $K=\left(c/4\pi f r_0\right)^2r_0^{\alpha}$.}
		\label{fig:LSvsFreq}
	\end{figure}

	As an example, we consider the case in which we use the path loss law $\ell(r) = K r^{-\alpha}$.  We select $K=\left(\frac{c}{4\pi f r_0}\right)^2r_0^{\alpha}$, where $f$ is the signal frequency and $r_0$ is a reference distance that we set to $c/2f$ m, corresponding to the far-field distance of a half-wavelength dipole antenna \cite{Balanis_book_2005}.  This $K$ corresponds to a free-space path loss \cite{Goldsmith_Wireless_Communications_2005}.  For this example, we choose $A$ to be a \Adb~dB difference, and the parameters $\mu=0$ and $\sigma=4$ which corresponds to an outdoor setting \cite{Goldsmith_Wireless_Communications_2005,Berg_et_al_1992}.  Plotting \eqref{eq:NoLambdaT} as a function of $\beta$ for a frequency of $f=1$~GHz gives us \figref{fig:LSvsBeta_NoLambdaT}.  Similarly, plotting  \eqref{eq:NoLambdaT} as a function of frequency $f$ for $\beta=0.95$ gives us \figref{fig:LSvsFreq}.
	
	As a practical example, let us consider deploying RadioHound sensors across South Bend, IN, which has an area of 41.88 square miles (108.47 km$^2$).  Assuming the city experiences a path loss of approximately $\alpha = 3$, we see from \figref{fig:LSvsBeta_NoLambdaT} that to have a 90\% confidence that we have sufficiently detected each source, we would need about 300 sensors/km$^2$.  Consequently, we would need on the order of 30,000 RadioHound sensors to have a 90\% confidence that the sources in South Bend are sufficiently detected.
	
	\section{Future Work}
	\subsection{Deployment Density}
	To further refine the estimate of sensor density, we can consider the effects of fading, which we expect will increase the required density.  Fading can be treated by having the radius of coverage be a random process of the angle (with some smoothness conditions).  To make the problem more tractable, we can make use of our former strategy by dividing the coverage region into sectors and calculating the average radius for that sector.  From there, we can calculate the average coverage area and apply the Boolean model.
	
	Additionally, we may consider different regimes of operation for our system.  In particular, we may consider the case of unique coverage in which a sensor covers a source uniquely and leverage the results of \cite{Haenggi_Sarkar_2016}.  This would tell us how much higher the sensor density would need to be in order to resolve the locations of closely-spaced sources.
	
	On the other hand, we may consider the density needed to form a percolation with high probability \cite[Ch.~9]{Haenggi_Stog}.  In this regime, we have a lower sensor density but form a giant connected component of coverage.  The benefit of such a large component of coverage is that a traveling source would not be able to evade detection when traveling from north to south or from east to west across the region.  
	
	\subsection{RF Power Map}  \label{subsec:Future Work System Performance}
	
	One open question surrounding the RF power map is how to obtain the $z$ value for the triples used to generate the map.  On one front, we need to determine the relevant value to compute from a single recorded periodogram.  Currently, we integrate the periodogram over the specified frequency range to obtain a power measurement.  Nonetheless, another choice may be to use the mean, median, or maximum value of the periodogram within the frequency range, which would enable comparison across different bandwidths of specified frequencies.  
	It is not immediately clear which option should be implemented. On a second front, it is possible that multiple measurements exist for a given location and time range.  Consequently, we must fuse multiple records into a single value.  Once again, we could take the average, maximum, or median value, each with potential benefits and drawbacks.
	
	Next, we can improve the interpolation method used to generate the RF power maps.  One of the problems with planar interpolation is that the interpolation values can never be larger than the maximum value of the sampled data.  In other words, $\max f(x,y) \le \max z$, which is inconsistent with standard path-loss models, for example.
	
	To overcome this drawback, we could use radial basis functions to model the sources and perform interpolation via a feed-forward neural network such as that used in \cite{Chen_et_al_1991}.  The problem then becomes where to place the sources.
	
	We propose using an iterative method to help place the sources until the resulting interpolant converges and describes the sampled data well enough according to some threshold on the MSE.  The first iteration could be planar interpolation to help determine where radial basis functions should be placed initially.  Then several iterations of the neural network could be performed to adjust the locations and/or number of sources until convergence is achieved.  The drawback to such an approach is the computational complexity and the amount of time required to generate an RF power map.  A potential benefit is that we may not need as high of a deployment density of sensors since the interpolation method could infer the presence of a far-away source by placing a radial basis function in the corresponding location.
	
	\section*{Acknowledgment}
	
	The authors thank Martin Haenggi for his thoughtful feedback regarding our use of stochastic geometry.
	
	\bibliographystyle{IEEEtran}
	
	\bibliography{IEEEabrv,Kleber_DySPAN_2017_bib}
	
	
\end{document}